\documentclass{article}
\usepackage{arxiv}
\usepackage[utf8]{inputenc}
\usepackage[T1]{fontenc}
\usepackage{hyperref}
\usepackage{booktabs}
\usepackage{graphicx}
\usepackage{natbib}
\usepackage{multirow}
\usepackage{import}

\title{Evaluating Eye Movement Biometrics in Virtual Reality: A Comparative Analysis of VR Headset and High-End Eye-Tracker Collected Dataset}

\author{ {\hspace{1mm}Mehedi Hasan Raju}\thanks{corresponding author} \\
    Texas State University\\
        601 University Drive\\
        San Marcos, Texas, 78640, USA\\
    \texttt{m.raju@txstate.edu} \\
    \And
    {\hspace{1mm}Dillon J. Lohr} \\
        Texas State University\\
        601 University Drive\\
        San Marcos, Texas, 78640, USA\\
    \texttt{djl70@txstate.edu} \\
    \And
    {\hspace{1mm}Oleg V. Komogortsev} \\
        Texas State University\\
        601 University Drive\\
        San Marcos, Texas, 78640, USA\\
    \texttt{ok@txstate.edu} \\
}

\date{}

\renewcommand{\shorttitle}{Evaluating Eye Movement Biometrics in Virtual Reality}

\begin{document}
\maketitle

\begin{abstract}

Previous studies have shown that eye movement data recorded at 1000 Hz can be used to authenticate individuals. 
This study explores the effectiveness of eye movement-based biometrics (EMB) by utilizing data from an eye-tracking (ET)-enabled virtual reality (VR) headset (GazeBaseVR) and compares it to the performance using data from a high-end eye tracker (GazeBase) that has been downsampled to 250~Hz. 
The research also aims to assess the biometric potential of both binocular and monocular eye movement data. 
GazeBaseVR dataset achieves an equal error rate (EER) of 1.67\% and a false rejection rate (FRR) at $10^{-4}$ false acceptance rate (FAR) of 22.73\% in a binocular configuration. 
This study underscores the biometric viability of data obtained from eye-tracking-enabled VR headset.

\end{abstract}


\section{Introduction}

We recognize that personal identity is not just defined by visible biological markers such as facial features, fingerprints, and iris configurations, as widely noted in biometric research \citep{jain2007handbook,biometric2010}, but also by subtle behavioral traits, including handwriting, vocal patterns, and eye movements.

For the last couple of decades, there has been an increasing curiosity in using eye movements for biometric analysis \citep{Kasprowski2004, abdelwahab2022deep}. 
Distinctive and complex eye movement patterns offer a range of benefits for biometric systems, including user identification \citep{schroder2020robustness, rigas2017current, deepeyedentification, deepeyedentificationlive} and authentication \citep{lohr2022eye, lohr2020metric, Lohr2020, lohrTBIOM,raju2023importance}. These patterns provide high specificity in individual recognition \citep{bargary2017individual} and are useful in detecting various disorders \citep{nilsson2016screening, billeci2017integrated}. Additionally, eye movements can predict demographic attributes like gender \citep{sargezeh2019gender, al2020gender}, and offer security features such as resistance to impersonation \citep{eberz2015preventing, Komogortsev2015} and capabilities for liveness detection \citep{rigas2015, raju2022iris}.

This study uses the latest technology in Eye-Tracking-enabled Virtual Reality (ET-enabled VR) headsets. 
It aims to understand the complexities of personal identity by analyzing eye movement patterns.  This field is noted for its intricacies and significant potential in biometric authentication. 
Our research is navigated through data characteristically noisy, inherent to ET-enabled VR headsets. 

In this report, we will evaluate the efficacy of EMB authentication through data obtained from an ET-enabled VR headset, operating at a sampling rate of 250 Hz. We will be using a monocular eye movement as primary data to work with. We will compare the biometric performance between binocular and monocular studies. We will also compare the biometric performance between the ET-enabled VR-collected dataset and a high-end eye-tracker collected dataset.

\section{Prior Work}

The pioneering work of Kasprowski and Ober in 2004 introduced the use of eye movements as a biometric modality for authenticating individuals, marking a significant advancement in the field \citep{Kasprowski2004}. This development triggered a wave of research into eye movement biometrics, as detailed in various studies \citep{brasil2020eye, zhang2015survey, zhang2016biometrics}, focusing on advancing state-of-the-art methodologies for user authentication based on eye movements. 

Present studies in biometric authentication frequently center on eye movement data \citep{deepeyedentificationlive, lohr2022eye, lohr2020metric, Lohr2020, lohrTBIOM, deepeyedentification}, with these movements demonstrating resistance to imitation or spoofing \citep{rigas2015,raju2022iris, Komogortsev2015}. 
The use of machine learning, especially deep learning, has increased within the field of eye movement biometrics. This includes two prevalent strategies: one that processes pre-extracted features \citep{lohr2020metric, george2016score} and another that derives embeddings directly from the raw eye-tracking data \citep{Jia2018, lohrTBIOM, deepeyedentification, deepeyedentificationlive, abdelwahab2022deep, lohr2022eye}. 

In recent years much research has been conducted demonstrating eye movement biometrics in virtual reality \citep{lohr2018implementation, lohr2020eye, lohr2023demonstrating}. 
Our goal is to expand this research to data acquired through an ET-enabled VR headset, operating at a 250 Hz sampling rate and known for its higher noise levels in data capture, in comparison to an eye-tracker like the EyeLink 1000.

\section{Methodology}

\begin{figure}[htbp]
\centering
\includegraphics[width=0.7\textwidth]{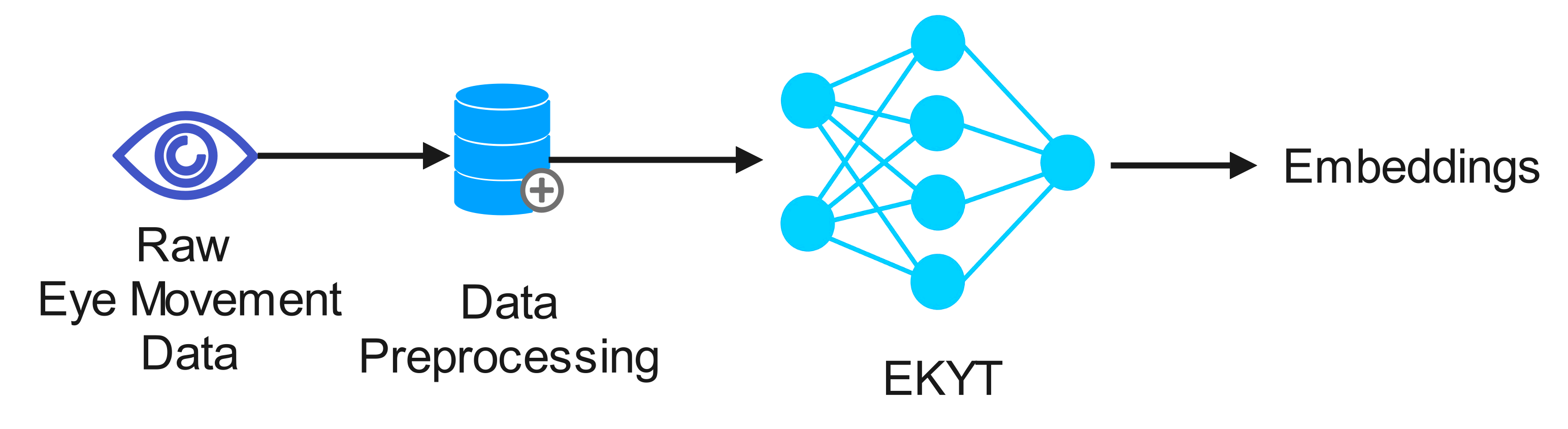}
\caption{Simple block diagram of the methodology.}
\label{fig:methodology}
\end{figure}

Fig.~\ref{fig:methodology} presents the block diagram of the methodology. 
The process begins with raw data, followed by a series of data preprocessing steps to remove inconsistencies. Finally, the processed data is fed into the Eye Know You Too (EKYT) model \citep{lohr2022eye}, which generates embeddings—a numerical representation that can be used in eye-movement-based biometric authentication \citep{lohr2022eye, raju2023importance, raju2024analysis}.
We are utilizing the EKYT network architecture, training procedure, and evaluation methodology, which, to our knowledge, represents the state-of-the-art (SOTA) in publicly available, end-to-end, machine learning-driven architectures. We direct readers to \citep{lohr2022eye} for more comprehensive methodological details.

\subsection{Dataset}
The primary dataset is GazeBaseVR (GBVR) \citep{lohr2023gazebasevr}, collected with an ET-enabled VR headset.
It includes 5020 binocular recordings from a diverse population of 407 college-aged subjects. 
The data was collected over 26 months in three rounds (Round 1 to Round 3). 
All the eye movements were recorded at 250~Hz sampling rate.
Each recording captures both horizontal and vertical movements of both eyes in degrees of visual angle (dva). 
Each participant completed a series of five eye movement tasks: vergence (VRG), horizontal smooth pursuit (PUR), reading  (TEX), video-viewing (VD), and a random saccade task (RAN).
For the GBVR dataset, the term ``short-term'' refers to the data collected in Round 1 with approximately 20-minute intervals between sessions, and the term ``long-term'' refers to the data collected in Round 3, which was obtained with approximately 26-month intervals from Round 1.
More details about the dataset and how data were collected are available in \citep{lohr2023gazebasevr}.

The compared dataset, we used in this study is the publicly available GazeBase (GB) dataset \citep{griffith2021gazebase}.
Eye movement recordings of this dataset are collected with a high-end eye-tracker, EyeLink 1000 at a sampling rate of 1000~Hz.
It includes 12,334 monocular recordings (left eye only) from 322 college-aged subjects.
The data was collected over three years in nine rounds (Round 1 to Round 9). 
Each recording captures both horizontal and vertical movements of the left eye in dva. 
Participants completed seven eye movement tasks: random saccades (RAN), reading (TEX), fixation (FXS), horizontal saccades (HSS), two video viewing tasks (VD1 and VD2), and a video-gaming task (Balura game, BLG). 
Each round comprised two recording sessions of the same tasks per subject, spaced by 20 minutes.
For the GB dataset, the term ``short-term'' refers to the data collected in Round 1 with approximately 20-minute intervals between sessions, and the term ``long-term'' refers to the data collected in Round 6, which was obtained with approximately a year interval from Round 1.
Further details about the dataset and recording procedure are available in \citep{griffith2021gazebase}.

\begin{figure}[htbp]
\centering
\includegraphics[width=0.85\textwidth]{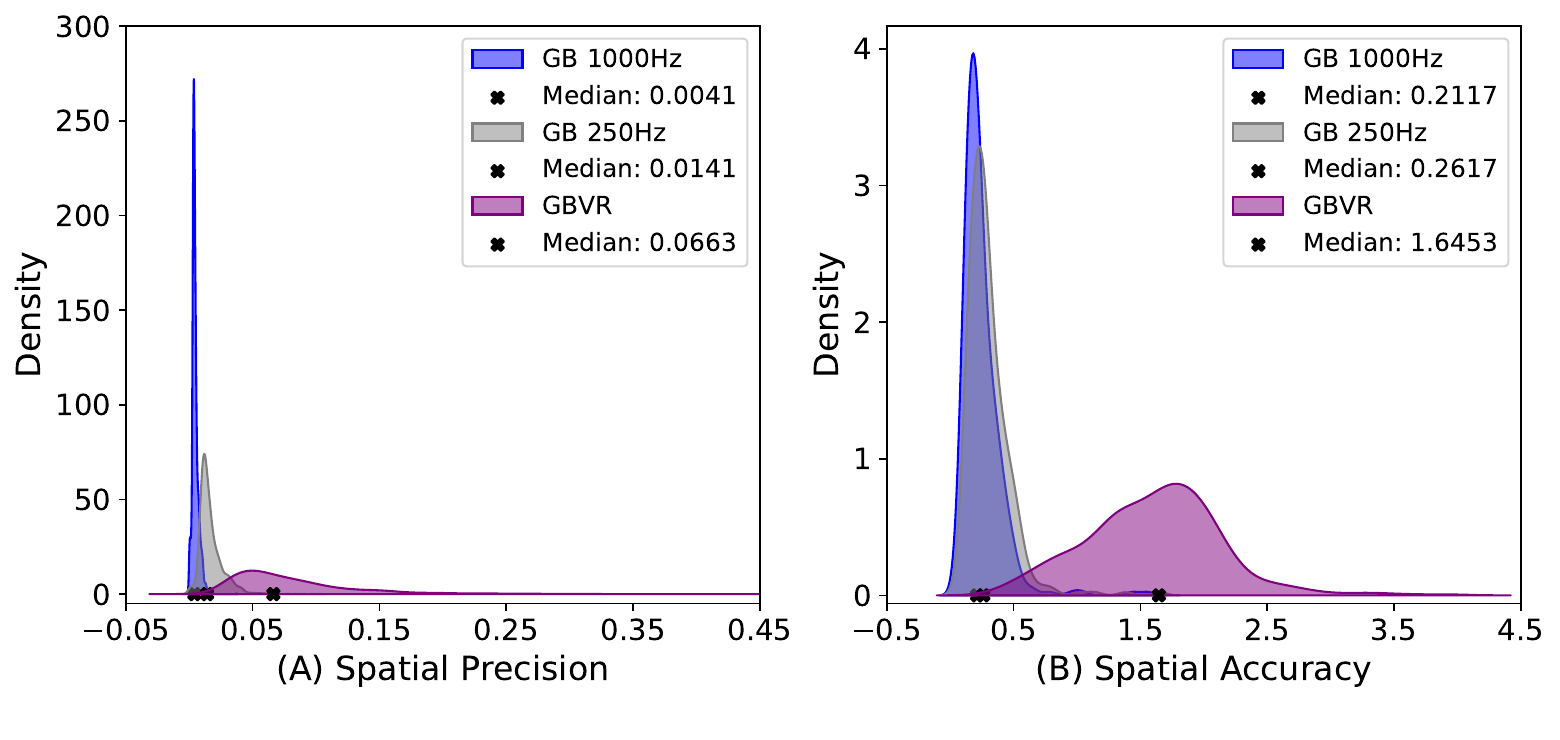}
\caption{Comparison of Kernel Density Estimations of Spatial Precision and Accuracy Measurements across datasets: GB, GB-250Hz (downsampled), and GBVR. 
(A) Illustrates the distribution of spatial precision measurements highlighting their respective medians. 
(B) Displays the distribution of spatial accuracy, also indicating median values. The plots demonstrate the variability within each dataset and visually compare the central tendencies and distribution shapes between the datasets.}
\label{fig:distribution}
\end{figure}

A direct comparison between GB and GBVR is challenging because they were collected at different sampling frequencies and differ in spatial accuracy and precision. To facilitate a fair comparison, we downsampled GB to 250Hz (termed GB-250Hz in the paper). Despite these adjustments, significant differences remain (as seen in Fig.~\ref{fig:distribution})\footnote{Spatial precision and accuracy were calculated using the approach suggested in \citep{lohr2019evaluating}.}. 
Readers should consider these factors when interpreting the findings presented in this paper.

\subsection{Data Preprocessing}
GBVR dataset was collected at a sampling rate of 250 Hz. The timestamps were supposed to be 4 milliseconds (ms) apart. However, this regularity wasn't maintained in the raw data. We applied linear interpolation to ensure consistent 4 ms intervals between timestamps.
In the GB dataset, EyeLink 1000 measures horizontal gaze positions between -23.3 and +23.3 dva, and vertical positions between -18.5 and +11.7 dva. Gaze samples outside the screen dimensions were also set to Not a number (NaN).
We processed the raw eye movement signals to create two velocity channels
\footnote{For the GBVR monocular study, we utilized data from the left eye movements whereas, in the binocular study, data from both the left and right eyes were used, expanding the input samples to four channels: left-horizontal and left-vertical velocity, and right-horizontal and right-vertical velocity}, 
horizontal and vertical, employing the Savitzky-Golay filter \citep{savitzkyGolayM} with a window size of 7 and an order of 2 \citep{friedman2017method}.
Velocity values were clamped between ±1000°/s to reduce the impact of noise on the data.
This method effectively smooths the input data for further analysis. 
Subsequently, we applied a rolling window technique to segment the recordings into non-overlapping 5-second windows, each containing 1250 samples for the GBVR dataset and 5000 samples for the GB dataset. 
For a comprehensive evaluation, these 5-second segments were aggregated into larger 45-second segment, with nine of these shorter segments forming each 45-second segment.

\subsection{Eye Know You Too network Architecture}

We have employed the Eye Know You Too (EKYT) network, a DenseNet-based architecture for eye-movement-based biometric authentication, as detailed in \citep{lohr2022eye}. This network achieves exceptional performance with high-quality GazeBase dataset, reaching a 0.58\% Equal Error Rate (EER) during reading-based authentication using 60 seconds of eye movement data. EKYT features eight convolutional layers, with each layer’s feature maps concatenated with those from preceding layers, enhancing feature extraction. The network then flattens these maps, and processes them through a global average pooling layer and a fully connected layer, resulting in a 128-dimensional embedding. For further details on EKYT's architecture, refer to \citep{lohr2022eye}.

\subsubsection{Dataset Split}

For the GBVR dataset, our training methodology for the model utilized data from the first three rounds of GBVR dataset. Round 1 included 407 subjects, while Round 3 involved 50 participants. Notably, the 50 subjects from Round 3 were originally part of the Round 1 subjects. To ensure the robustness of our validation process, we isolated the data from these 50 individuals from both Round 1 and Round 3, treating it as a held-out dataset. This data was not utilized in either the training or the validation phases.

For the GB dataset, the model was trained using data from Rounds 1-5, excluding the BLG task. Round 1 included 322 subjects, and Round 6 included 59 subjects, all of whom were also part of the Round 1. The data from these 59 subjects across both rounds were set aside as a held-out dataset and were not utilized in the model's training or validation processes.

For the cross-validation process, we divided the remaining pool of participants into four distinct, non-overlapping folds. This division aimed at achieving a balanced distribution of participants and their corresponding recordings across all folds. Each fold comprised unique classes, ensuring no overlap in data. The specifics of our fold assignment strategy, which plays a crucial role in the reliability of our cross-validation, are thoroughly detailed in \citep{lohrTBIOM}.

\subsubsection{Training Procedure and Evaluation}
We trained four distinct models, each employing one unique validation set (a held-out fold) and the remaining three folds as the training set. We utilized the Adam optimizer\citep{adam} and PyTorch's OneCycleLR\citep{Lr} with cosine annealing for managing the learning rate. The training was based on the multi-similarity loss (MS)\citep{Wang2019}, using default hyperparameters as recommended by EKYT.
Each model was trained using input samples that featured two channels (horizontal and vertical velocity) over 1250 time steps; for the GazeBase dataset, this was extended to 5,000 time steps. Training spanned 100 epochs, starting with a learning rate of $10^{-4}$, which peaked at $10^{-2}$ by the 30th epoch, before gradually decreasing to $10^{-7}$ over the subsequent 70 epochs. The batch size was fixed at 64 samples, divided into 8 classes per batch, with each class containing 8 samples.
Biometric performance was evaluated over two distinct periods: short-term and long-term. 
Four models trained with 4-fold cross-validation generated 128-dimensional embeddings per window, concatenated into 512-dimensional embeddings, combining the models into an ensemble model.

All models were trained on a workstation equipped with an NVIDIA RTX A6000 GPU, an AMD Ryzen Threadripper PRO 5975WX with 32 cores, and 48 GB of RAM. The system ran an Anaconda environment with Python 3.7.11, PyTorch 1.10.0, Torchvision 0.11.0, Torchaudio 0.10.0, Cudatoolkit 11.3.1, and Pytorch Metric Learning (PML)~\citep{Musgrave2020a} version 0.9.99.

\subsection{Performance Metrics}
We assessed our model using three metrics:
\begin{itemize}
    \item Equal Error Rate (EER) is the point on a Receiver Operating Characteristic (ROC) curve where the rates of false acceptance (FAR) and false rejection (FRR) are equal \citep{rigas2015}. Lower EER signifies better system performance.

    \item Decidability index (d-prime) is a measure of the separation between genuine and imposter similarity distributions \citep{daugman2000biometric}. A larger d-prime indicates a clearer distinction between the genuine and imposter groups, suggesting better performance

    \item False Rejection Rate (FRR) \citep{conrad2015cissp} is at which authorized users are mistakenly rejected. A low FRR is desirable, but it should be balanced against False Acceptance Rate (FAR). In this study we have used FRR at FAR of $10^{-4}$, aiming to stay below 5\% suggested by FIDO standards \citep{FIDO2020}.

\end{itemize}

\section{Results}

\begin{table*}[ht]
\centering
\caption{Biometric performance of GBVR (monocular study) dataset in short-term and long-term evaluation for all the tasks. Arrows indicate whether a larger or smaller value is better.}
\begin{tabular}{|cl|c|c|c|c|}
\hline
\multicolumn{2}{|c|}{Round} & Task & EER (\%) $\downarrow$ & d-prime $\uparrow$ & FRR @ $10^{-4}$ FAR (±STD) \%  $\downarrow$\\ \hline
\multicolumn{2}{|c|}{\multirow{5}{*}{Short-term (Round-1)}} & TEX & 2.01 & 3.50 &18.04 (0.99) \\ \cline{3-6} 
\multicolumn{2}{|c|}{}  &   RAN  & 3.39     & 3.09     & 78.34 (11.07)      \\ \cline{3-6} 
\multicolumn{2}{|c|}{}  &   PUR  & 7.57     & 2.60     & 93.74 (4.22)       \\ \cline{3-6} 
\multicolumn{2}{|c|}{}  &   VD   & 8.31     & 2.72     & 81.52 (5.28)       \\ \cline{3-6}
\multicolumn{2}{|c|}{}  &   VRG  & 13.33    & 2.24     & 91.66 (0.19)       \\ \hline
\multicolumn{2}{|c|}{\multirow{5}{*}{ Long-term (Round-3)}} & TEX   &  11.75 &  2.40 &  73.17 (0.70) \\ \cline{3-6} 
\multicolumn{2}{|c|}{}  & RAN  & 18.33   & 1.82     & 89.53 (1.46)   \\ \cline{3-6} 
\multicolumn{2}{|c|}{}  & PUR  & 25.00   & 1.29     & 94.89 (0.44)   \\ \cline{3-6} 
\multicolumn{2}{|c|}{}  & VD   & 20.00   & 1.65     & 94.72 (0.94)   \\ \cline{3-6}
\multicolumn{2}{|c|}{}  & VRG  & 30.00   & 1.03     & 96.54 (0.46)   \\ \hline
\end{tabular}
\label{table:GBVRresults}
\end{table*}

Table~\ref{table:GBVRresults} presents the results on the GBVR dataset.
It provides biometric performance on the GBVR dataset (monocular) in short-term and long-term evaluations for all the tasks.
The performance metrics used were EER, d-prime, and FRR @ $10^{-4}$ FAR.
In the short-term, the TEX task has the lowest EER (2.01\%), the highest d-prime (3.50), and the lowest FRR @ $10^{-4}$ FAR  of 18.04\% which is the best-performing result on GBVR dataset.
For the long-term evaluation, all the performance metrics are worse than the R1 reading task. The detailed results on the GBVR dataset are given in Table~\ref{table:GBVRresults}.

\begin{table*}[ht!]
\centering
\caption{Comparison of biometric performance for GBVR to GB and GB-250Hz. Arrows indicate whether a larger or smaller value is better. Performance from the reading task (TEX) has been stated here.}
\begin{tabular}{|cc|c|c|c|}
\hline
\multicolumn{2}{|c|}{Metrics $\rightarrow$} & {EER (\%)} $\downarrow$ &  {d-prime} $\uparrow$ &  {FRR @ $10^{-4}$ FAR (±STD) \% $\downarrow$ } \\ \hline
\multicolumn{1}{|c|}{\multirow{4}{*}{Short Term (R1)}} & GBVR (Monocular) & 2.01 & 3.50 & 18.04 (0.99) \\ \cline{2-5} 
\multicolumn{1}{|c|}{} & GBVR (Binocular)   & 1.67 & 3.73 & 22.73 (2.41) \\ \cline{2-5} 
\multicolumn{1}{|c|}{} & GB-250Hz   & 4.50 & 3.64 & 11.67 (0.83) \\ \cline{2-5} 
\multicolumn{1}{|c|}{} & GB  & 0.41 & 4.88 & 5.07 (0.16) \\ \hline
\multicolumn{1}{|c|}{\multirow{2}{*}{Long Term}}  & GBVR (Monocular)  & 11.75 & 2.40 & 73.17 (0.70) \\ \cline{2-5} 
\multicolumn{1}{|c|}{\multirow{2}{*}{(GBVR: R3 and GB: R6)}}  & GBVR (Binocular) & 10.25 & 2.58 & 89.05 (8.73) \\ \cline{2-5} 
\multicolumn{1}{|c|}{} & GB-250Hz & 10.17 & 2.68 & 64.19 (0.93) \\ \cline{2-5} 
\multicolumn{1}{|c|}{} & GB  & 5.08 & 3.76 & 23.58 (0.56) \\ \hline
\end{tabular}
\label{table:comparison}
\end{table*}

In Table~\ref{table:comparison}, we have compared the biometric performance of the GBVR dataset in monocular and binocular settings with GB and GB-250Hz datasets. Performance from the reading task is mentioned in the table because of two main reasons: 1) Aligning with established practices in previous EMB research where the TEX task is used for comparison and 2) TEX shows the better result in Table~\ref{table:GBVRresults}.

In the short-term evaluation, GB demonstrates superior performance across all assessed metrics as expected. 
Specifically, it records an EER of 0.41\%, a d-prime of 4.88, and an FRR at a FAR of $10^{-4}$ is 5.07\%. 
GB-250Hz dataset exhibits an EER of 4.50\%, a d-prime of 3.64, and an FRR at $10^{-4}$ FAR of 11.67\%. 
In a monocular setup, GBVR shows an EER of 2.01\%, a d-prime of 3.50, and an FRR at $10^{-4}$ FAR of 18.04\%. 
Notably, the binocular study using GBVR surpasses both the monocular GBVR and GB-250Hz in terms of EER (1.67\%) and d-prime (3.73).

In long-term evaluations, GB consistently maintains the best performance across metrics. 
It achieves an EER of 5.08\%, a d-prime of 3.76, and an FRR at $10^{-4}$ FAR of 23.58\%. 
GB-250Hz reports an EER of 10.17\%, a d-prime of 2.68, and an FRR at $10^{-4}$ FAR of 64.19\%. 
In the monocular study, GBVR records an EER of 11.75\%, a d-prime of 2.40, and an FRR at $10^{-4}$ FAR of 73.17\%. 
The binocular study of GBVR achieves an EER of 10.25\%, a d-prime of 2.58, and an FRR at $10^{-4}$ FAR of 89.05\%.

\section{Discussion}

\subsection{Performance Comparison: GBVR (Monocular) vs GBVR (Binocular)}

\begin{figure}[htp]
    \centering
    \begin{minipage}[b]{0.45\linewidth}
        \includegraphics[width=\linewidth]{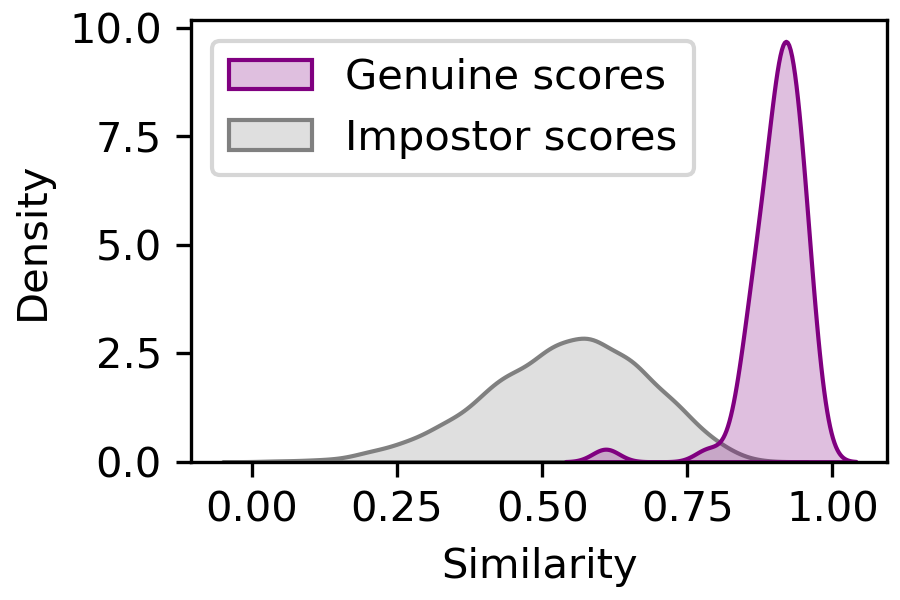}
    \end{minipage}
    \begin{minipage}[b]{0.45\linewidth}
        \includegraphics[width=\linewidth]{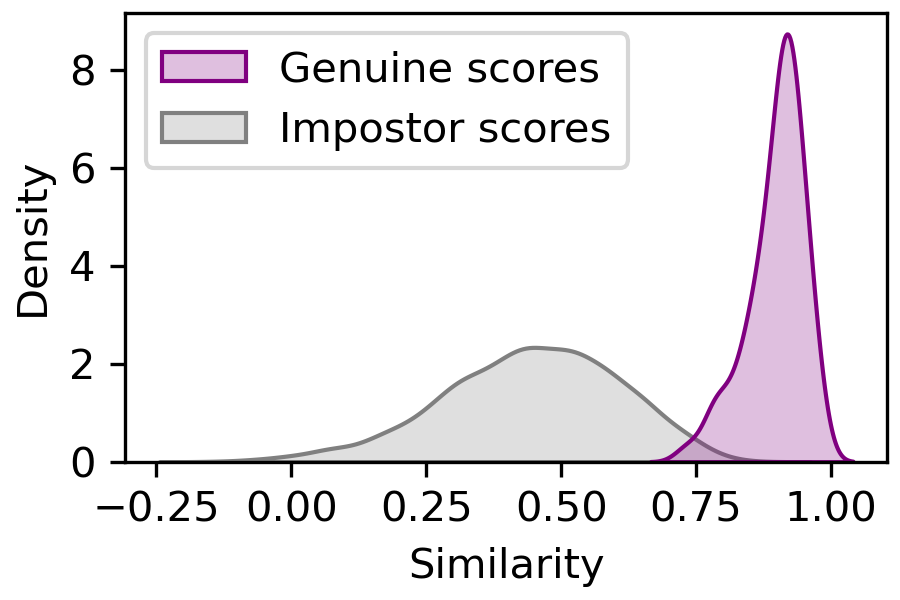}
    \end{minipage}
    \caption{Similarity score distribution for GBVR monocular study (left) and binocular study (right).}
    \label{fig:binovsmono}
\end{figure}

\begin{figure}[htp]
    \centering
    \begin{minipage}[b]{0.45\linewidth}
        \includegraphics[width=\linewidth]{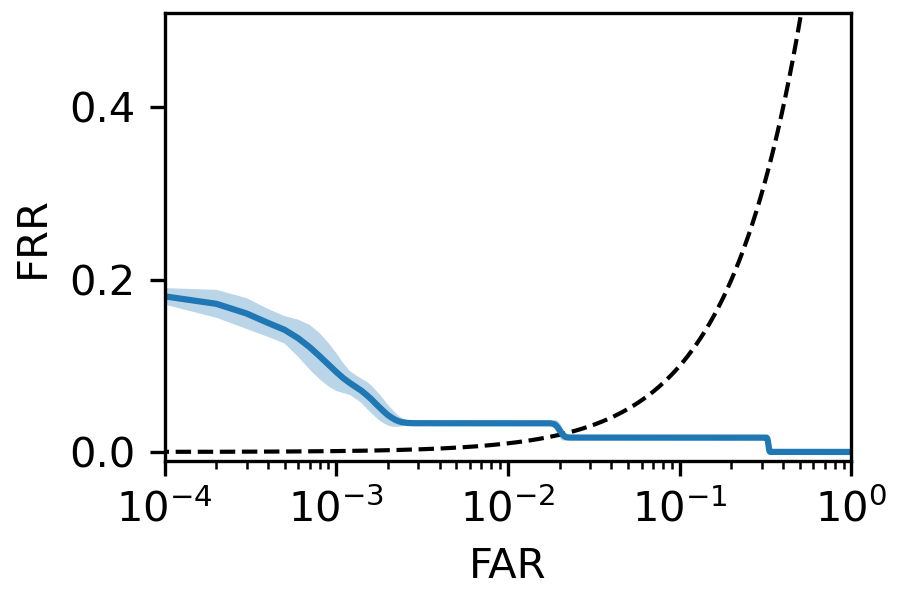}
    \end{minipage}
    \begin{minipage}[b]{0.45\linewidth}
        \includegraphics[width=\linewidth]{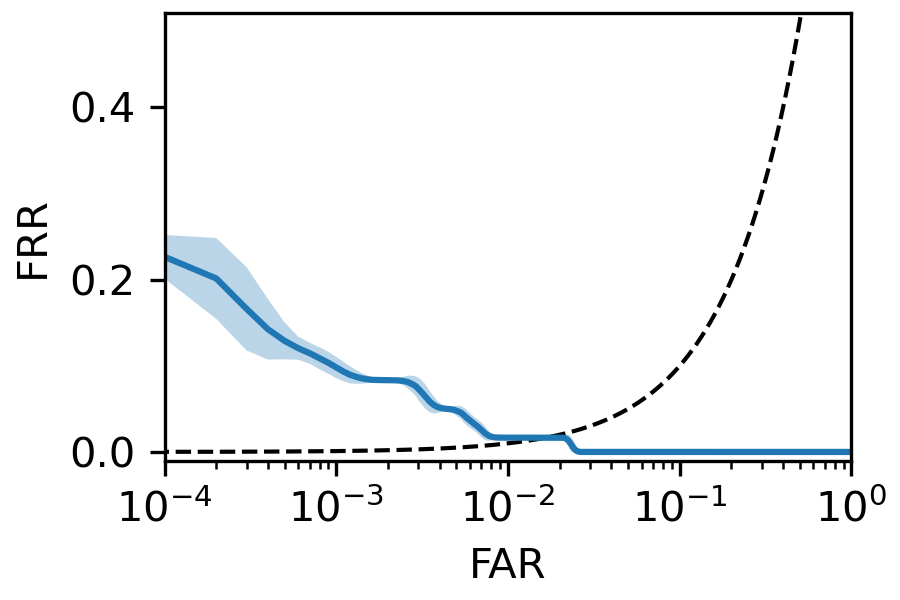}
    \end{minipage}
    \caption{ROC Curve for GBVR monocular study (left) and binocular study (right).}
    \label{fig:binovsmonoroc}
\end{figure}

In this research, we included both monocular and binocular studies of GBVR dataset.
For the GBVR monocular study, we utilized data from the left eye whereas, in the binocular study, data from both the left and right eyes were used. The binocular study shows a lower EER than the monocular study, which indicates it is more accurate in distinguishing between genuine users and impostors—there's a smaller chance for both false acceptances and rejections to occur at the point where their rates are equal. 
The d-prime, which measures the separability of genuine and impostor distributions, is also higher for the binocular study, suggesting a better separation between distributions, refer to Fig.~\ref{fig:binovsmono}. 
However, the FRR at $10^{-4}$ FAR is higher for the binocular study, which implies that it is more likely to reject a genuine user.
Refer to Fig.~\ref{fig:binovsmonoroc} for the ROC curve.
This pattern is valid for both short-term and long-term evaluation. 

The binocular study seems to offer better biometric performance due to a lower EER and higher d-prime both in the short term and long term, but at the cost of convenience as seen with a higher FRR, especially over the long term.

\subsection{Performance Comparison: GBVR (Binocular) vs GB-250Hz}

The primary focus of this paper is to present EMB performance in the authentication paradigm on data collected with an ET-enabled VR headset and compare it against that of a high-end eye tracker. 
It is to be noted that ET-enabled VR headset data are very noisy compared to the laboratory standard eye-tracker.

\begin{figure}[htp]
    \centering
    \begin{minipage}[b]{0.45\linewidth}
        \includegraphics[width=\linewidth]{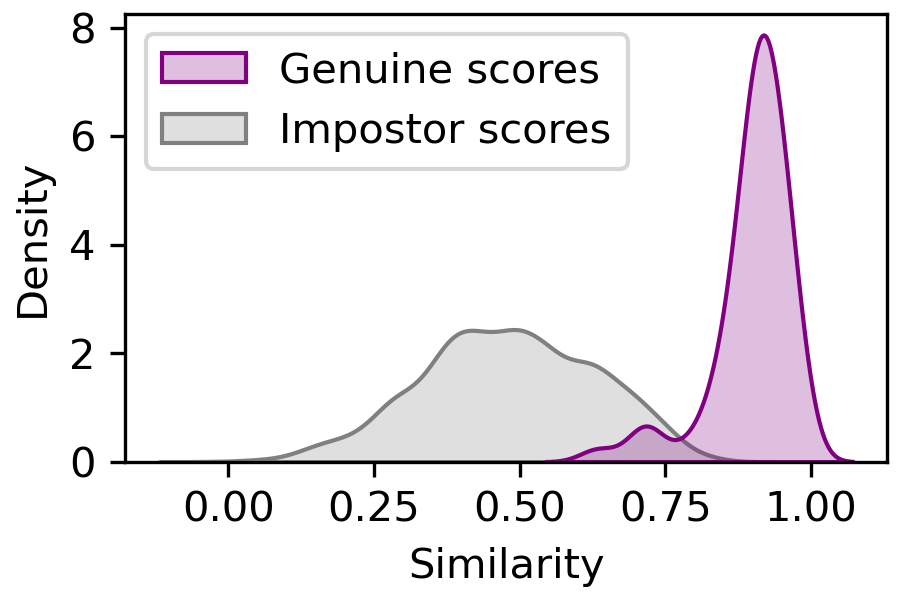}
    \end{minipage}
    \begin{minipage}[b]{0.45\linewidth}
        \includegraphics[width=\linewidth]{Figures/simdist_gbvr_binocular.png}
    \end{minipage}
    \caption{Similarity score distribution for GB-250Hz (left) and GBVR binocular study (right).}
    \label{fig:binovsgb250}
\end{figure}

\begin{figure}[htp]
    \centering
    \begin{minipage}[b]{0.45\linewidth}
        \includegraphics[width=\linewidth]{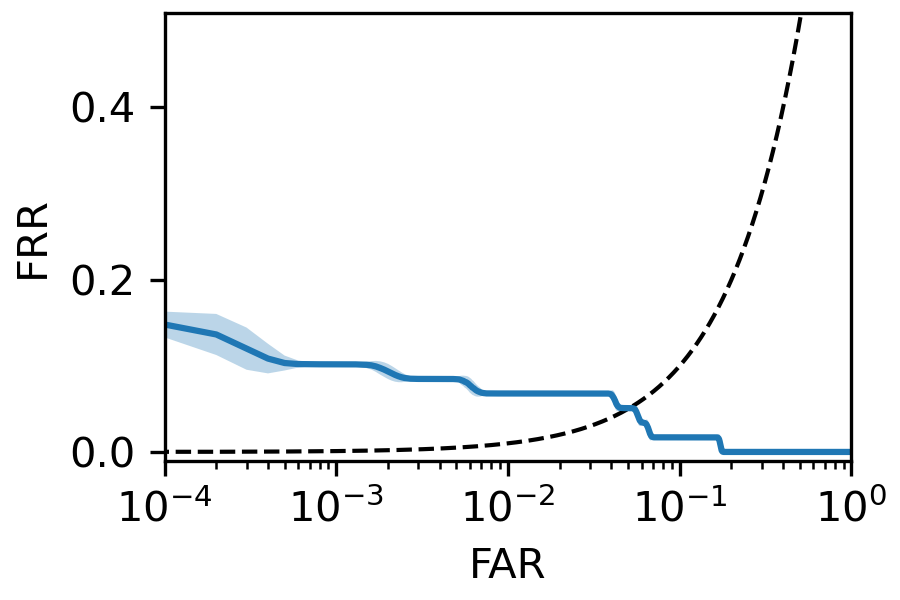}
    \end{minipage}
    \begin{minipage}[b]{0.45\linewidth}
        \includegraphics[width=\linewidth]{Figures/roc_gbvr_binocular.png}
    \end{minipage}
    \caption{ROC Curve for GB-250Hz (left) and GBVR binocular study (right).}
    \label{fig:binovsgb250roc}
\end{figure}

Yet, we have achieved an EER of 1.67\%, d-prime of 3.73, and an FRR at $10^{-4}$ FAR is 22.73\% (±2.41). for short-term evaluation.
On the other hand, GB-250Hz results in an EER of 4.50\%, d-prime of 3.64, and an FRR at $10^{-4}$ FAR of 11.67\% (±0.83).
Refer to Fig.~\ref{fig:binovsgb250roc} for the ROC curve.
Except for FRR, GBVR (binocular) outperforms GB-250Hz in both EER and d-prime performance metrics. 
Over the long term, the EER values for both analyses increased significantly, indicating that the biometric performance dropped over time. 
Similarity score distribution for short-term evaluation is shown in Fig.~\ref{fig:binovsgb250}. 
GB-250Hz has slightly lower EER and higher d-prime than GBVR (binocular), suggesting it performs slightly better over the long term. 
Moreover, the FRR is substantially lower for GB-250Hz than for the GBVR (binocular), with less variability (lower STD), indicating that GB-250Hz has a lower possibility of rejecting genuine users over time.

In two of three performance metrics, the GBVR (Binocular) dataset outperforms the GB-250Hz dataset in short-term evaluation, proving the underlying potential of ET-enabled VR headsets' recorded eye movements as a biometric authentication method. The performance especially in short-term evaluation is quite promising.  

\section{Conclusion}

In conclusion, this study demonstrates that bionocular eye-tracking data from a VR headset is highly effective for biometric authentication, achieving an Equal Error Rate (EER) of 1.67\% and d-prime of 3.73 outperforming the biometric performance of monocular GBVR data and GB-250Hz dataset in short-term evaluation. In terms of FRR @ $10^{-4}$ FAR is 22.73\% which is not the best but it shows promises.
Additionally, short-term data is shown to be more reliable for biometric authentication than long-term data. 
These findings suggest the advantage of binocular data in VR eye-tracking environments and the potential in eye-movement-based biometric authentication systems.

\bibliographystyle{unsrtnat}
\bibliography{ref}

\end{document}